%% file: main.tex
\documentclass[sigconf,10pt]{acmart}
\setcopyright{rightsretained}
\renewcommand\footnotetextcopyrightpermission[1]{}

\fancypagestyle{firstpage}{
  \fancyhf{} 
  \fancyfoot[C]{\thepage}
}

\newcommand{\showcomments}{0}

\usepackage{tikz}
\usepackage{amsmath}
\usepackage{graphicx}
\usepackage{hyperref} 

\usepackage{booktabs}
\usepackage{tabularx} 

\usepackage{color}
\usepackage{colortbl}

\usepackage{caption}
\usepackage{subfig}
\usepackage{algorithm,algpseudocode}

\usepackage{balance}

\usepackage{enumitem}

\usepackage{ifthen} 

\usepackage[T1]{fontenc}

\graphicspath{{./figures/}}

\begin{document}

\sloppy

\frenchspacing

\date{}


\title{Visualizing Distributed Traces in Aggregate} 


\author{Adrita Samanta$^{\star}$, Henry Han$^{\star}$, }
\author{Darby Huye$^{\dagger}$, Lan Liu$^{\dagger}$, Zhaoqi Zhang$^{\dagger}$, Raja R. Sambasivan$^{\dagger}$}
\affiliation{%
  \institution{$^{\star}$MIT PRIMES 2023, $^{\dagger}$Tufts University}
}

\input{std_commands}

\clubpenalty=150
\widowpenalty=150

\maketitle

\input{abstract}

\input{./sections/intro.tex}
\input{./sections/background.tex}
\input{./sections/goal.tex}
\input{./sections/methods.tex}
\input{./sections/visualization.tex}

\input{./sections/results.tex}
\input{./sections/future_work.tex}
\input{./sections/conclusion.tex}


\balance
\bibliographystyle{plain}
\bibliography{manual} 

\end{document}

%% file: std_commands.tex


\newcommand{\code}[1]{\texttt{\textbf{#1}}}

\newcommand{\term}[1]{\emph{#1}}


\newcommand{\minorsection}[1]{\textbf{#1}:}

\newcommand{\minicaption}[2]{\caption[#1]{\textbf{#1.} #2}}

\newcommand{\units}[2]{#1~#2}

\newcommand{\command}[1]{{\sc \MakeLowercase{#1}}}

\newcommand{\fix}[1]{\marginpar{\LARGE\ensuremath{\bullet}}
    \MakeUppercase{\textbf{[#1]}}}
\newcommand{\fixnote}[2]{\marginpar{\LARGE\ensuremath{\bullet}}
    {\textbf{[#1:} \textit{#2\,}\textbf{]}}}

\newcommand{\setmargins}[4]{
    \setlength\topmargin{#3}
    \addtolength\topmargin{-.5in}  
    \addtolength\topmargin{-\headheight}
    \addtolength\topmargin{-\headsep}
    \setlength\textheight{\paperheight}
    \addtolength\textheight{-#3}
    \addtolength\textheight{-#4}

    \setlength\oddsidemargin{#1}
    \addtolength\oddsidemargin{-1in}
    \setlength\evensidemargin{\oddsidemargin}
    \setlength\textwidth{\paperwidth}
    \addtolength\textwidth{-#1}
    \addtolength\textwidth{-#2}
}

\newcolumntype{L}{X}
\newcolumntype{C}{>{\centering\arraybackslash}X}
\newcolumntype{R}{>{\raggedleft\arraybackslash}X}

\long\def\omitit#1{}

\newcommand{\inlinesection}[1]{\smallskip\noindent{\textbf{#1.}}}

\newenvironment{outlineenv}{\par\color{blue}}{\par}
\newenvironment{pagelenenv}{\par\color{red}}{\par}

\newcommand{\outline}[1]{\begin{outlineenv}#1\end{outlineenv}}
\newcommand{\pagelenblah}[1]{\begin{pagelenenv}#1\end{pagelenenv}}

\newcommand{\pagelen}[1]{
   \ifthenelse{\equal{\showcomments}{1}}{
     \pagelenblah{#1}}{}}

\newcommand{\outlinetext}[1]{
   \ifthenelse{\equal{\showcomments}{1}}{
     \outline{#1}}{}}

\newcounter{todoctr}
\newcounter{authoractr}
\newcounter{authorbctr}
\newcounter{authorcctr}
\newcounter{reviewerctr}
    
\newcommand{\todo}[1]{%
     \ifthenelse{\equal{\showcomments}{1}}{
	\stepcounter{todoctr}%
	\textcolor{red}{\textbf{TODO~\arabic{todoctr}: #1}}     
     }
     {} 
}
\newcommand{\darby}[1]{%
    \ifthenelse{\equal{\showcomments}{1}}{
	\stepcounter{authoractr}%
	\textcolor{orange}{Darby~\arabic{authoractr}: #1}
    }
    {} 
}
\newcommand{\raja}[1]{%
     \ifthenelse{\equal{\showcomments}{1}}{
	\stepcounter{authorbctr}%
	\textcolor{blue}{Raja~\arabic{authorbctr}: #1}%
     }
     {} 
}
\newcommand{\maxliu}[1]{%
     \ifthenelse{\equal{\showcomments}{1}}{
	\stepcounter{authorcctr}%
	\textcolor{green}{Max~\arabic{authorcctr}: #1}%
     }
     {} 
}

\newcommand{\reviewer}[2]{%
     \ifthenelse{\equal{\showcomments}{1}}{
	\stepcounter{reviewerctr}%
	\textcolor{green}{reviewer #1: ~\arabic{reviewerctr}: #2}%
     }
     {} 
}

\newcounter{missingctr}
\newcommand{\missingval}[1]{\stepcounter{missingctr}\underline{\textbf{MISSING\_VAL~\arabic{missingctr}}}}

\newcommand{\algoname}{Casper}
\newcommand{\casper}{Casper}
\newcommand{\Casper}{Casper}

\newcommand{\rpcid}{\texttt{rpcid}}
\newcommand{\rpctype}{\texttt{rpctype}}
\newcommand{\um}{\texttt{UM}}
\newcommand{\dm}{\texttt{DM}}
\newcommand{\rt}{\texttt{rt}}
\newcommand{\traceid}{\texttt{traceid}}

%% file: abstract.tex
\textbf{Abstract} \\
\quad Distributed systems are comprised of many components that communicate together to form an application. Distributed tracing gives us visibility into these complex interactions, but it can be difficult to reason about the system's behavior, even with traces. Systems collect large amounts of tracing data even with low sampling rates. Even when there are patterns in the system, it is often difficult to detect similarities in traces since current tools mainly allow developers to visualize individual traces. Debugging and system optimization is difficult for developers without an understanding of the whole trace dataset. In order to help present these similarities, this paper proposes a method to aggregate traces in a way that groups together and visualizes similar traces. We do so by assigning a few traces that are representative of each set. We suggest that traces can be grouped based on how many services they share, how many levels the graph has, how structurally similar they are, or how close their latencies are. We also develop an aggregate trace data structure as a way to comprehensively visualize these groups and a method for filtering out incomplete traces if a more complete version of the trace exists. The unique traces of each group are especially useful to developers for troubleshooting \cite{sifter}. Overall, our approach allows for a more efficient method of analyzing system behavior.

%% file: sections/intro.tex
\section{Introduction}
\quad Distributed systems are characterized by the utilization of and communication between multiple independent components on different machines to complete actions. Their monolithic counterparts are built as a single unit that contains all components and services as a single executable. As programmers realize the scalability and performance advantages over monoliths, distributed systems have become increasingly prevalent in today's era of networking. 

Distributed tracing, a technique used in modern-day applications, is a critical part of logging and debugging these applications. Tracing is a method of monitoring and analyzing the behavior of large and complex systems like microservices by "tracing" individual calls and requests as they travel across the different services. Tracing provides insights into the interactions of different parts of a distributed system and measures the performance of requests. It helps developers identify latency issues and problems in the system. However, software companies can generate up to millions of traces daily (or more!), and combing through all of them can be inefficient for investigating performance issues. Thus, it can be difficult to analyze which parts of the system are causing issues. Going through many common traces is a huge waste of time and effort \cite{sifter}, so one of our goals is to create an algorithm to find and group similar traces together to make debugging easier. This would enable programmers to more easily analyze and debug large quantities of traces. Additionally, we desire to create a method for visualizing the entire group of similar traces without just looking at one trace that represents the group. It is also important for developers to see the entire workflow to understand how each of the services is connected. However, traces can sometimes be incomplete due to issues like memory overflow. We want to be able to filter out incomplete traces from the trace dataset if a more complete version of the trace exists in the set.

%% file: sections/background.tex
\section{Background}
\quad
Companies collect millions of traces for developers to examine, comprised of operations performed throughout the execution of a request. \darby{revise the following sentence, it's unclear what you  mean-- error traces and requests?? } If they gather millions of traces as well as requests, it would be impossible to know which traces to look at first to find the behaviors that cause errors. Thus, tools are necessary to help developers identify paths that show unexpected behaviors. 

We introduce some examples of the most common methods of distributed tracing that are used to analyze systems, and we explain the flaws of these current techniques. We also give context to the sources that inspired our approach, including the ideas of Jaccardian Similarity and Disjoint Set Union.

\darby{it'd be cool if you named your tool-- not necessary but fun!}
Our research uses these concepts to create a tool to group together similar traces based on a trace similarity definition. We then select one trace per group to represent the group. Additionally, we present an aggregate trace graph that captures additional information that cannot be expressed in a single representative trace in each group. 
We also present a method for filtering out incomplete traces so that we have a smaller and more valuable trace set to analyze.

\subsection{Models for Traces}
\quad Distributed tracing can be implemented in multiple approaches. One of the most common types is span-based tracing. 

Span-based tracing consists of a tree of spans, where each span represents a single operation from its time of initiation to its end. Each parent span can create multiple child spans. For example, a child span could be the authentication of a user that is called from a client request, and the request would represent the parent span. All of the spans are then combined to form a trace diagram, which represents the interactions between all of the services utilized. These dependencies between spans are represented by edges and operate based on caller-callee relationships. These are useful for capturing timing information and the duration of specific operations. Programmers can use this information to detect services that have high latency or performance errors.

One common method to visualize components of all traces in a system is by using a dependency diagram. Because it contains information throughout the entire trace dataset, dependency diagrams aren't necessarily representative of any individual trace. It uses distinct nodes to represent different services as well as edges to show communication between nodes, allowing developers to see common interactions. 

\subsection{Issues with Current Tracing Methods}
\quad When a distributed tracing system is set up for a company, the developers are faced with analyzing millions of traces to find the sources of performance issues and bugs in their applications, such as latency issues. Moreover, it can even be a challenge to just understand the system from only a trace dataset. Analyzing every single trace can be very time-consuming, especially when many traces are nearly identical, so developers try to only analyze the unique traces to help with troubleshooting \cite{sampleHST}. Since manually going through a whole dataset to find interesting traces for debugging can be difficult, we created an algorithm that can group similar traces together into tracing groups that also allows for a better understanding of a system. We then use these groups of similar traces and select one trace to represent the group which we call representative traces. These representative traces are like the unique traces that the developer would find if they were to go through all traces manually.

We also want to be able to create a visualization for a data structure that can represent an entire group of similar traces without just looking at a single trace.

Sometimes, individual traces can be missing data (missing services or requests). This is possible when spans are lost when they are supposed to be stored or when a service that is processing chooses to not complete the trace. We want to filter out incomplete traces so that we analyze complete tracing data.

\subsection{Prior Work}
\quad Previous work has been done with the goal of reducing the time and effort needed for going through many distributed traces. One such tool is tprof \cite{tprof}, a performance profiler that creates an aggregate trace by taking the average of subspan durations and timings. For example, in Figure~\ref{fig:tprofexample}, if a span A was recorded with latencies of 4ms, 5ms, and 6ms, the aggregate trace would depict a duration of 5ms for span A. 

\input{figures/fig_tprof_example}

However, one downside of this definition is that it doesn't give the whole picture of a group of traces. Since it only shows the average of all spans, developers won't be able to see the variation in the group, which is important to gain a better sense of aggregate traces. Instead, it would be more beneficial to incorporate a diagram that shows the full spread of all the different trace groups. tprof is also difficult to scale significantly since it separates different orderings of spans into different groups, which becomes impossible when considering thousands of workflows with thousands of services. Additionally, the data would be overwhelming for developers, as tprof would need to create innumerable groups to accommodate for each ordering of each trace.

It has recently been shown that traces are often missing data \cite{meta}. This missing data is the cause of either a lack of instrumentation or just missing services in traces due to errors in the storage of the spans or services not completing the process of adding information into the trace. Currently, not much work has been done on fixing the incomplete traces to create the proper system workflow.

One tool that is used to aggregate trace data is Zeno\cite{zeno}. Zeno uses temporal provenance, the ability to record the history of changes to data over time, to detect performance issues in distributed systems and dependencies. It uses this information to construct an aggregate structure with three categories of services: a backbone of requests, sources of delays, and services without delay. Some services are depicted in the groups to give a better understanding of each category. Edges are formed between services to show common dependencies. This graph presents developers with an easy way to find sources of delays. However, temporal provenance does not always determine the root causes for delay from an unusual factor since it does not know what is unusual. Another drawback is that small but unrelated factors that add up will not show a root cause.

Another graph that clusters traces is presented in SampleHST \cite{sampleHST}. Its two axes are labeled $m$ and $p$. $m$ represents the mass score of a trace--the level of deviation from a standard behavior. Higher mass scores correlate to higher chance of anomalies. $p$ shows the percentile value of the mass of a trace in comparison to all other traces. Thus, a smaller $p$ indicates that a trace's mass score is higher than fewer traces, which makes it less anomalous. Traces are clustered based on these axes, which gives a good understanding of how many potentially anomalous traces there are and how much they deviate. However, its sampling strategy allows for some anomalous traces to be left out. \darby{should add a sentence to this paragraph and zeno one above with how their approaches are lacking still (motivating our work)}

\subsection{Jaccardian Similarity} 
\quad The Jaccardian Similarity \cite{jaccard-similarity} between two objects is used to measure their similarity. To find the Jaccard Index between sets, it is calculated by dividing the intersection between the sets by the union of them. In other words: 

{
\begin{figure}[h]
    \centering
    \vspace{-0.3cm}
    \includegraphics[width=3cm]{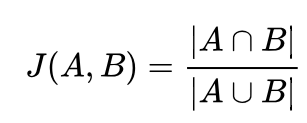}
    \vspace{-0.5cm}
\end{figure}
}

It is commonly used in data science applications such as data mining and E-commerce. We use Jaccardian Similarity to measure the similarity between two sets that each describe a trace. We use this similarity to decide whether the traces are similar or not.

\subsection{Disjoint Set Union} 
\quad The Disjoint Set Union algorithm \cite{disjoint-set-union} is used to efficiently track collections of disjoint sets. It allows sets to be grouped together through the usage of representatives. The process begins with many groups of individual sets. Within each group, a representative set stands for the other sets in the group. Whenever two groups are merged, one of the two representative sets is chosen as the new representative, and it is applied to all of the newly joined sets. Repeating this process allows for the creation of a few large groups that are organized by their representative set. If two sets have the same representative, they are in the same group. However, if they have different representatives, they are in different groups. 

%% file: figures/fig_tprof_example.tex
{
\begin{figure}[t]
    \includegraphics[width=0.9\columnwidth]{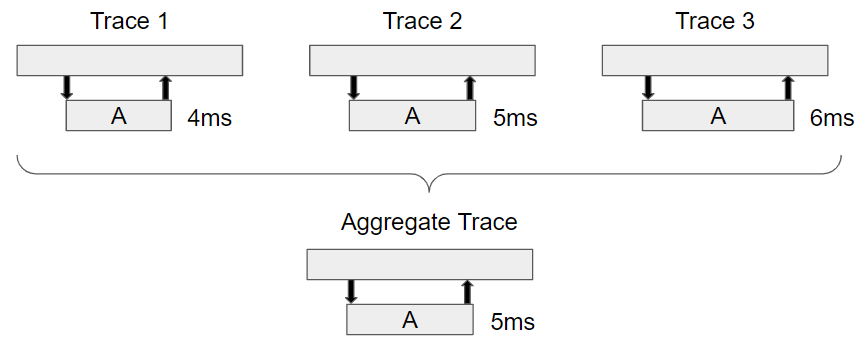}
    \minicaption{Tprof Aggregate trace of 3 traces}{}
    \label{fig:tprofexample}
    \vspace{-0.5cm}
  \end{figure}
}

%% file: sections/goal.tex
\section{Research Questions}
One of our biggest concerns is how we can group traces together by similarity. With traces organized into similar groups, we can then:
\begin{itemize}
    \item Filter out the set of traces so that we only keep track of and analyze the "most complete" version of every incomplete trace. We use this processed and smaller set of traces for the rest of our analysis and visualization.
    \item Group similar traces together and select a trace that is the most similar to the other traces in its corresponding group (a trace that "represents" the group). We call this selected trace a "representative trace". 
    \item Create an aggregate trace data structure to visualize the groups of representative traces in a comprehensive manner.
\end{itemize} 

\subsection{Preprocessing Traces}
\quad We filter out incomplete traces from our data set so that developers only analyze the most complete traces. Developers often analyze a set of millions of traces, so we assume that every incomplete trace must have a “completed version” of itself. So, if a trace has a "more complete" version of itself in the set, we remove the incomplete version. We keep removing the incomplete version of traces so that at the end we have a smaller set of traces that still provides the same information as the original set of traces. 

\subsection{Identifying Groups of Similar Traces}
\quad To group similar traces together, we first determine possible definitions for trace similarity and define a threshold value to be able to group the traces into a reasonable number of trace groups. We also want to choose one representative trace for each group.

\subsection{Aggregate Trace Data Structure}
\quad We also aim to depict the groups of similar traces in a way that is comprehensive for developers. We designed an aggregate trace data structure to organize trace data from distributed systems.
\darby{update this text-- depicting groups of representative traces isn't exactly what you're doing-- you're taking groups of similar traces and displaying all variation in the group in an aggregate data structure. }

In order to properly depict traces and groups, we would want to use a graph data structure. We need to find the best way to visualize the data from our definitions of trace similarity.

%% file: sections/methods.tex
\section{Preprocessing Traces/Grouping Similar Traces}
We group similar traces as shown in Figure~\ref{fig:design_diagram}. First, we \textit{preprocess traces} to address our first research goal (as discussed in \S~\ref{sec:similar:preprocessing}). \S~\ref{sec:similar:groups:encoding} discusses how we \textit{encode traces} for comparison based on various definitions of similarity. \S~\ref{sec:similar:groups:applydef} discusses how we \textit{build a similarity graph} which connects traces with similar workflows from which we can \textit{identify similar traces}. Once we have groups of similar traces, we can select a \textit{representative trace} from each group, which addresses our second research goal. We discuss visualization methods in \S~\ref{sec:visualize}.

\input{figures/fig_design_diagram}

\subsection{Preprocessing Traces}
\label{sec:similar:preprocessing}
We aim to filter out incomplete versions of traces to have a smaller and more valuable set of traces to analyze. In this section, we discuss our approach to identifying incomplete traces and their corresponding completed version. 

\textbf{Identifying Incomplete-Complete Trace Pairs} 
We defined an incomplete trace to be a subgraph of a more complete version of the trace. For example, in Figure~\ref{fig:preprocessing_traces} we have that trace 1, an incomplete trace, has a more completed version of itself in the trace set, which would be trace 2 because trace 1 is a subgraph of trace 2.
In order to filter out all incomplete traces from the set, we follow these steps:
\begin{enumerate}
    \item Extract the traces from the JSON files.
    \item Store each trace as the set of their edges.
    \item For each pair of traces, check if one trace is the subgraph of another (the set of edges of one trace would be a subset of the set of edges of the other trace). If the trace is a subgraph of the other trace, that means we have found an incomplete-complete trace pair. So, we remove the incomplete trace from the tracing set.
\end{enumerate}

Figure~\ref{fig:preprocessing_design_diagram} is a flowchart to show the major steps of our preprocessing method in a more understandable manner.

\input{figures/fig_preprocessing_traces}

\input{figures/fig_preprocessing_design_diagram}

\subsection{Identifying Groups of Similar Traces}
\label{sec:similar:groups}
We aim to take a set of traces and group similar traces together. In this section, we explore multiple potential definitions for trace similarity. We then apply the concepts of Jaccardian Similarity and Disjoint Set Union to form our distinct groups of traces. We also define thresholds for checking if two traces are similar enough to be in the same trace group.

\subsubsection{Encoding Traces} 
\label{sec:similar:groups:encoding}
We first looked at different ways of representing/encoding traces:

\textbf{Definition 1} (services in the trace) 

\textbf{Details:} In this definition, we want to keep track of the names of the services in the trace (which are the nodes in the trace for our model of trace representation). A representative trace, using this definition, will represent all traces with the same list of services in its trace.

\textbf{Uses:} With the representative traces and their corresponding groups, we can help developers fix bugs in the application. For example, we may have a trace composed of the services "Authentication" and "User Credential Cache". This request could be logging a user in or verifying it has the correct credentials to perform a requested action. We group together traces that share the same services since they are likely to be doing similar types of actions. 

\textbf{Definition 2} (number of levels in the trace) 

\textbf{Details:} In this definition, we want to keep track of the depth of the graph. The depth of the graph is the maximum number of nodes it takes to go from a root node to a leaf node. A representative trace, using this definition, will represent all traces with the same depth in their trace graph. 

\textbf{Uses:} We realized that there are many cases where two traces have the same number of levels in the graph but are actually completely different and not really comparable. We also notice that this definition won't be very useful because knowing the depth of a trace graph doesn't help developers. The depth of a trace is a structural definition that won't help developers fix bugs in their application or optimize it. 

\textbf{Definition 3} (exact structure of the trace graph) 

\textbf{Details:} In this definition, we want to keep track of the exact structure of the trace graph by using its edge list. A representative trace represents all traces with similar graph structure. 

\textbf{Uses:} With the representative traces and their corresponding groups, developers can try to use knowledge of bugs in traces to help fix bugs in traces with similar structure. This is because bugs might be similar in both traces since the graphs have the same structure. Let's look at an example where a developer gets two traces that were put into the same trace group by using this definition. Suppose that in one of the traces, there is a timeout error in a front-end application that makes requests to multiple databases, and in the other trace, there is also a timeout error from another type of application that also makes requests to databases. Since both traces have a timeout when an application makes multiple requests to databases, the developer might be able to use the same solution to the request problem in the systems. 

\textbf{Definition 4} (similar latencies) 

\textbf{Details:} In this definition, we want to keep track of the latency when a service makes all of its calls. A representative trace will represent all traces with similar latency values. We categorize the latency values into ranges for fast, medium, and slow based on how fast the calls were made (fast, medium, and slow will be a range of latencies). 

\textbf{Uses:} Initially we thought that grouping together traces with similar latencies is helpful in finding common issues but we realized that measuring the latency to find representative traces is not effective. Many services can be in the same latency category even if they don't have any of the same properties which means that the groups found with this definition won't help developers. 

\subsubsection{Applying Chosen Definition to Find Similar Groups} 
\label{sec:similar:groups:applydef}
We used \textbf{Definitions 1 and 3} as our methods of encoding the traces. To apply these definitions we followed a set of steps:  
\begin{enumerate}
    \item Extract the traces from the JSON files.
    \item Apply the encoding method on each trace.
    \item Check how similar the traces are by comparing the encodings of each trace. Note that the encoding for each trace will be an array for both definitions so we measure the similarity between them using Jaccardian Similarity.
    \item If the traces are similar enough (i.e. the similarity measure passes some given threshold), we add an edge between their traces in a trace similarity graph to represent that the traces are similar. We set the threshold to be 0.8 for our initial tests but we discuss our method for finding an optimal threshold in \S~\ref{sec:similar:groups:similaritythreshold}.
    \item Apply Disjoint Set Union (DSU) to find the connected components (in this case, groups of similar traces) in the trace similar graph. 
    \item Look at the trace that has the highest degree in each group (meaning that it is the most similar to the other traces in the group). This trace of the highest degree will be the representative trace for each of the groups.
\end{enumerate}

\subsubsection{Similarity Threshold} 
\label{sec:similar:groups:similaritythreshold}
Something else to consider is how high or low the threshold for similarity should be. For example, consider a set of two traces called Trace 1 and Trace 2 as shown in figure~\ref{fig:example_traces}. Trace 1 contains nodes Front End, Friends Database, Post, Friends, Feed, while Trace 2 contains nodes Front End, Friends, Friends Database. 


\input{figures/fig_example_traces}

We calculate the similarity to be $\frac{3}{5}$ since the traces share 3 nodes out of the 5 total nodes between them. If our similarity threshold is less than or equal to $\frac{3}{5}$, we would draw an edge between them to represent a pair of similar traces and put both in the same group as shown in Figure~\ref{fig:example_group_1}.

\input{figures/fig_example_group_1}

On the other hand, if the threshold is larger than $\frac{3}{5}$, then the traces would not be connected by an edge and would not be considered similar. Trace 1 and 2 would each have their own separate group as shown in figure~\ref{fig:example_group_2}. 

\input{figures/fig_example_group_2}

Depending on how many groups we want to create, we should apply a different similarity threshold. For example, if we use a set of traces that share many services, a low threshold would aggregate all of them in the same group. A higher threshold would expose the differences between the traces by separating them into more groups. However, too high of a threshold would result in too many groups to analyze. Thus, we need to be able to vary the number of trace groups, so that developers can adapt to different trace datasets. It would allow multiple perspectives on a system and aid in understanding the most common types of traces and their performance issues.

\textbf{Choosing an Optimal Threshold}

For our initial tests, we used a threshold value of 0.8 which seemed to work for most cases. To find a threshold that is "optimal" for the given set of traces, we first set the parameter to represent the number of trace groups that is the goal (the "optimal" threshold should create the number of trace groups that is closest to this goal). We also implemented the methods discussed in \S~\ref{sec:similar:groups:applydef} a single function called $getGroups$ which has an input of $threshold$.

Then we defined procedures for each step: 
\begin{itemize}
    \item Step 1: Compare two thresholds to check which one is more "optimal"
    \item Step 2: Keep looking for two thresholds to compare based on previous information (i.e. the two thresholds compared should be more optimal than the thresholds that were compared previously). 
\end{itemize} 

Method for \textbf{Step 1}: We define the following variables: 

$threshA$ - threshold to compare 

$threshB$ - another threshold to compare 

$goalGroups$ - goal for the number of trace groups

We apply the methods defined in sections 4.2 and 4.3 to find similar groups for each of the thresholds (we run $getGroups(threshA)$ and $getGroups(threshB)$). As output of each run, we get the trace groups and we keep track of the number of groups for each threshold. Then, we compare the difference between the number of groups from each run to the goal number of groups ($goalGroups$). Whichever threshold gives a trace partition with the number of traces closest to the goal will "move to the next level", meaning it will be considered more optimal than the previous thresholds. 

Method for \textbf{Step 2}: We use a Binary Search approach to find the thresholds to compare. Note that we still use the $numGroups$ variable from Step 1. The threshold must always be in the range of 0 to 1 so we initialize $low = 0$ and $high = 1$. Now, we look at the midpoint (call it $mid$) between $low$ and $high$, which is initially $0.5$. We run $getGroups(mid)$ and compare the number of trace groups from this function run (which we can call $curGroups$) to $goalGroups$. 

If $curGroups$ is greater than $goalGroups$, then we have too many groups which means that we don't have enough edges in the trace similarity graph. If this is the case, we set $high = mid$ because the threshold must be smaller than $mid$. 

If $curGroups$ is smaller than $goalGroups$, then we don't have enough groups which means that we have too many edges in the trace similarity graph. If this is the case, we set $low = mid$ because the threshold must be larger than $mid$.

We repeat this process until our range is so small that we keep looking at the same values. Then we just output the current best threshold that we have seen.  

To get the final result, we run $getGroups$ on this optimal threshold that we have found.

%% file: figures/fig_design_diagram.tex
{
\begin{figure}[h]
    \includegraphics[width=0.8\columnwidth]{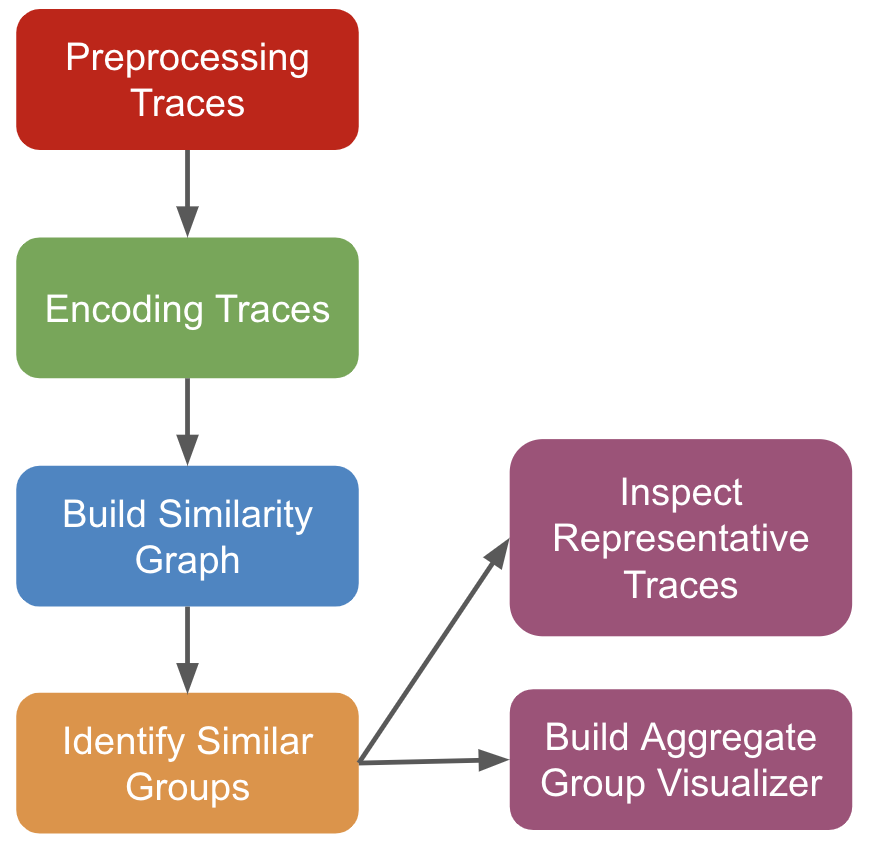}
    \minicaption{Design Diagram}{Flowchart displaying the major steps of our method.}
    \label{fig:design_diagram}
    \vspace{-0.5cm}
  \end{figure}
}

%% file: figures/fig_preprocessing_traces.tex
{
\begin{figure}[h!]
    \includegraphics[width=0.9\columnwidth]{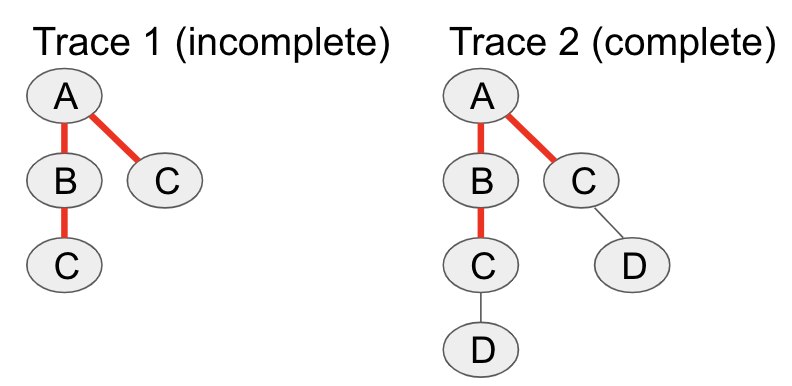}
    \minicaption{Preprocessing traces}{Example of two traces, trace 1 is the incomplete version of trace 2. The edges marked in red are the edges that are in both traces 1 and 2. Note: all edges in trace 1 are marked in red.}
    \label{fig:preprocessing_traces}
    \vspace{-0.5cm}
  \end{figure}
}

%% file: figures/fig_preprocessing_design_diagram.tex
{
\begin{figure}[h!]
    \includegraphics[width=0.9\columnwidth]{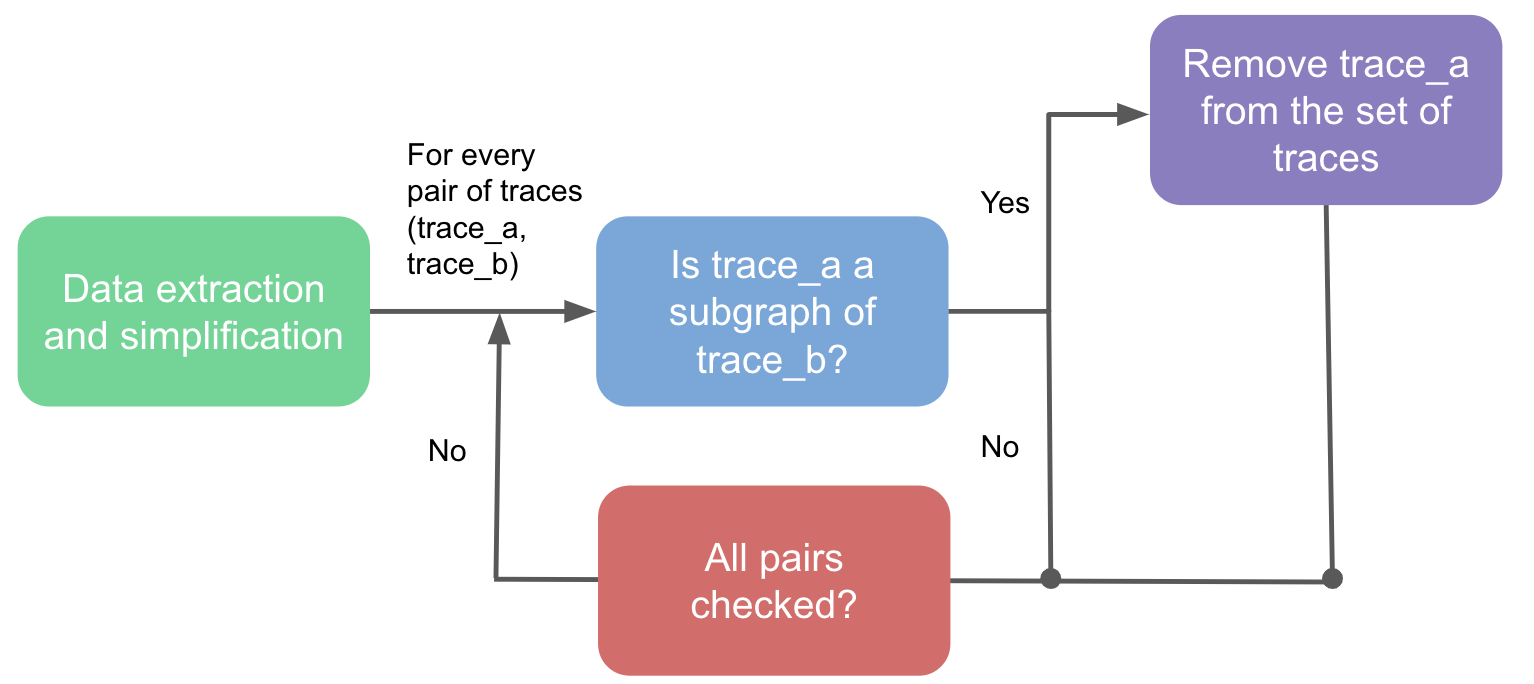}
    \minicaption{Preprocessing design diagram}{Flowchart displaying major steps of our preprocessing method which is discussed in \S~\ref{sec:similar:preprocessing}.}
    \label{fig:preprocessing_design_diagram}
    \vspace{-0.5cm}
  \end{figure}
}

%% file: figures/fig_example_traces.tex
{
\begin{figure}[t]
    \includegraphics[width=0.9\columnwidth]{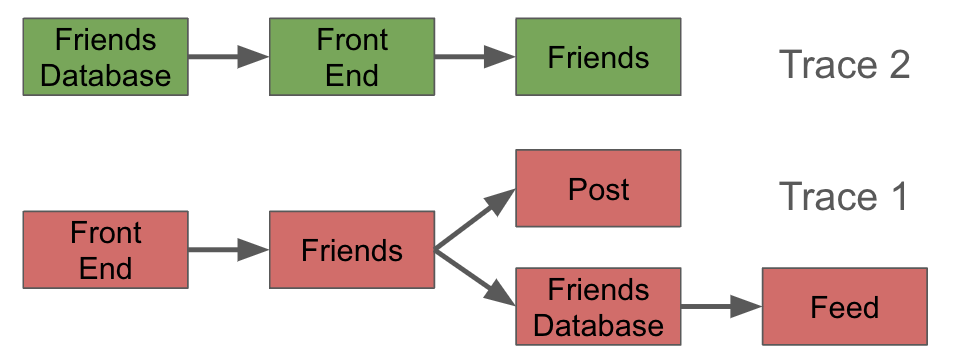}
    \minicaption{Example traces 1 and 2 with service names}{}
    \label{fig:example_traces}
    \vspace{-0.5cm}
  \end{figure}
}

%% file: figures/fig_example_group_1.tex
{
\begin{figure}[h!]
    \includegraphics[width=0.9\columnwidth]{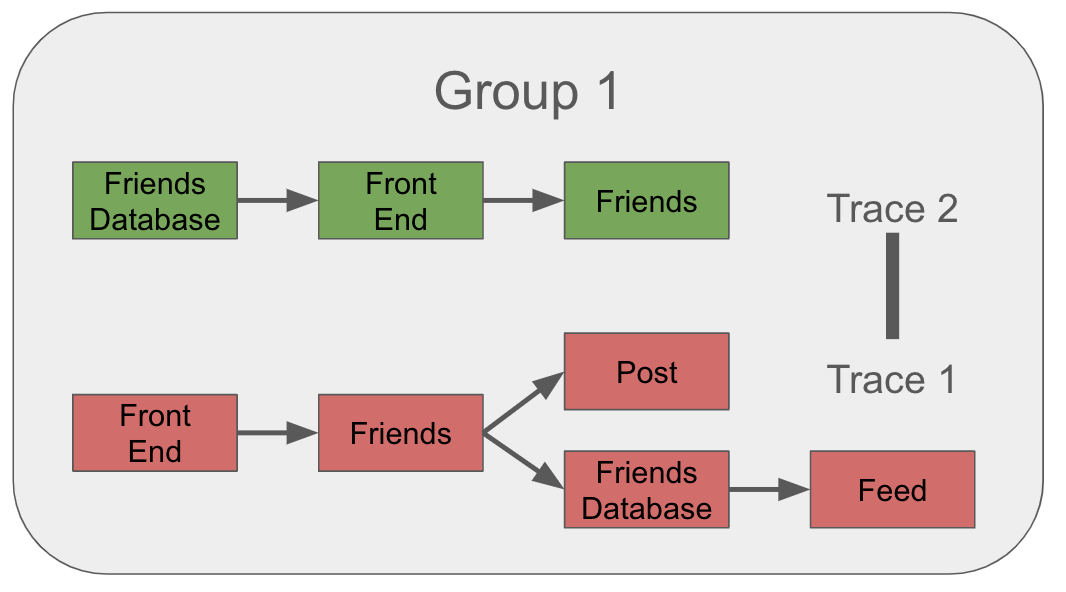}
    \minicaption{Group using similarity threshold under $\frac{3}{5}$}{}
    \label{fig:example_group_1}
    \vspace{-0.5cm}
  \end{figure}
}

%% file: figures/fig_example_group_2.tex
{
\begin{figure}[h!]
    \includegraphics[width=0.9\columnwidth]{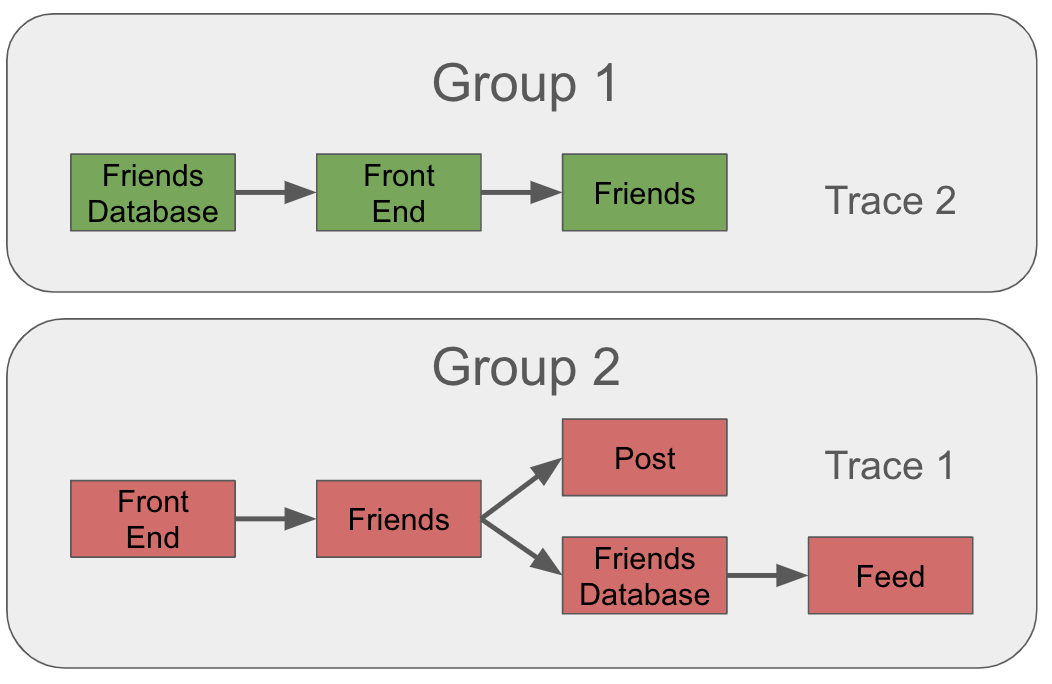}
    \minicaption{Group using similarity threshold above $\frac{3}{5}$}{}
    \label{fig:example_group_2}
    \vspace{-0.5cm}
  \end{figure}
}

%% file: sections/visualization.tex
\section{Visualizing Groups of Traces}
\label{sec:visualize}
\darby{I would refactor this section to have the following 3 subsections: converting a representative to visualization, visualizing a single group, visualizing a chosen service}
Visualizing similar groups of traces is another way to gain a better understanding of a trace dataset. To do so, we use graph-tool\cite{peixoto_graph-tool_2014} to visualize a group of traces in a way that highlights common services and allows for users to look into certain nodes. Additionally, we provide a view of a chosen service within a group and present its most common calls, giving a better idea of service interactions within a group. 
\todo{cite graph tool}
\darby{is the last sentence accurate to what you did? It reads to me like you're saying we put distance between groups of traces to show how similar the groups are-- is this what you mean?}

\subsection{Converting Representative Output to Visualization Format}
\darby{title should be updated-- should be something like converting from group of traces to vis}

\quad We took the output from our representative traces, comprised of groups of traces and services of each trace, \darby{output is groups of similar traces as far as I remember!} and reorganized them to fit the inputs we used for the graph-tool visualizations. By creating a library of trace spans, we can find how many service calls were made and what services were used. This is useful for applications to our \textbf{Definition 1} of service sets. \darby{I would add a sentence somewhere saying this visualization method was designed for the first def of similarity -- service sets}

First, we installed graph-tool in colab with \href{https://colab.research.google.com/github/count0/colab-gt/blob/master/colab-gt.ipynb#scrollTo=6km1lWMF2kAm}{this code}. 

Using Group 1 from Figure \ref{fig:example_group_1} gives the following output in Figure \ref{fig:rep_output} and following library in Figure \ref{fig:rep_library}. \darby{update rep node definition-- not the current wording we're using}

\input{figures/fig_rep_output}

\input{figures/fig_rep_library}

\darby{Make sure to say what threshold was used, or at least what output was used-- is this for when both traces are in the same group or in their own groups? }
We go through each trace's services to create a set of all the different services, which in this case are ['Front End', 'Feed', 'Friends Database', 'Post', 'Friends']. Then, we find which services are present in each trace and add them up to get a set of the total number of traces that each node is in. The corresponding set to the group of these 2 example traces would be [2, 1, 2, 1, 2]. To compare these values, we also store the total number of traces in the group, which is 2. Finally, we also look for service interactions, and we create a set for each node that corresponds with the number of services it calls across the group. For example, the Friends service and set [0,0,2,1,0] means that Friends calls Friends Database 2 times and Post 1 time throughout the 2 traces. 

Putting it all together, running the 2 traces from Figure \ref{fig:example_traces} gives the output of: 


\input{figures/fig_vis_output}

\subsection{Visualizing a Single Group}

We view individual groups by using a set of numbers that corresponds to how many times each node shows up in a different trace. A node that is present in all traces is highlighted in yellow, while a node that only shows up in some traces is colored gray. Additionally, we scale the size of the nodes to clearly show the more common services in the group. For example, if by using a group consisting of Traces 1 and 2 from Figure \ref{fig:example_traces} \darby{use ref tag to the specific section, after refactoring to section 4 is done}, we have 2 traces with a total of 5 services: Front End, Feed, Friends Database, Post, Friends. Its corresponding set of nodes is then [2, 1, 2, 1, 2]. Because the services Front End, Friends Database, and Friends each show up in every trace, their nodes are yellow in Figure 11.

\input{figures/fig_example_group_vis}

\subsection{Visualizing a Chosen Service}

Additionally, it would be useful to see into the interactions of a specific node in a group. We enable this by allowing developers to choose any service $A$ and highlighting it in green, as well as a trace group to visualize the nodes that $A$ calls. Furthermore, we scale the sizes of the edges from $A$ based on how many times $A$ calls a specific node in a group. For example, selecting the \textit{Friends} service in the group reveals its connections to \textit{Post} and \textit{Friends Database} Figure~\ref{fig:example_friends_group}. We can see that \textit{Friends} calls \textit{Friends Database} more frequently than it calls \textit{Post} by the thickness of the edges.


\input{figures/fig_friends_group_vis}

%% file: figures/fig_rep_output.tex

{
\begin{figure}[t]
    \centering
    \includegraphics[width=0.8\columnwidth]{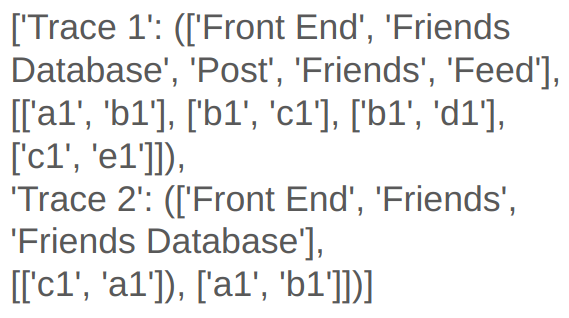}
    \minicaption{Representative output for traces 1 and 2}{}
    \label{fig:rep_output}
    \vspace{-0.5cm}
  \end{figure}
}

%% file: figures/fig_rep_library.tex
{
\begin{figure}[t]
    \centering
    \includegraphics[width=0.6\columnwidth]{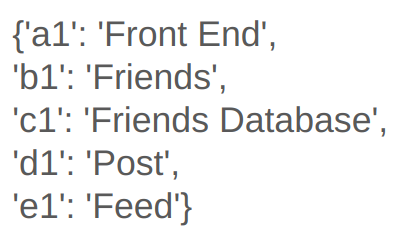}
    \minicaption{Span library for trace services}{}
    \label{fig:rep_library}
    \vspace{-0.5cm}
  \end{figure}
}

%% file: figures/fig_vis_output.tex
{
\begin{figure}[h]
    \centering
    \vspace{-0.4cm}
\includegraphics[width=0.8\columnwidth]{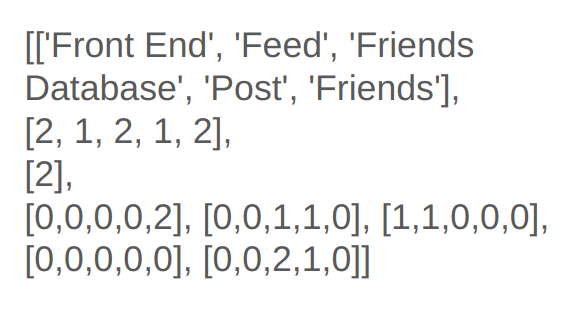}
        \vspace{-0.5cm}
\minicaption{Visualization output of node names, node occurrences, total traces, and calls per node}{}
    \label{fig:vis_output}
    \vspace{-0.2cm}
  \end{figure}
}

%% file: figures/fig_example_group_vis.tex

{
\begin{figure}[t]
    \includegraphics[width=0.9\columnwidth]{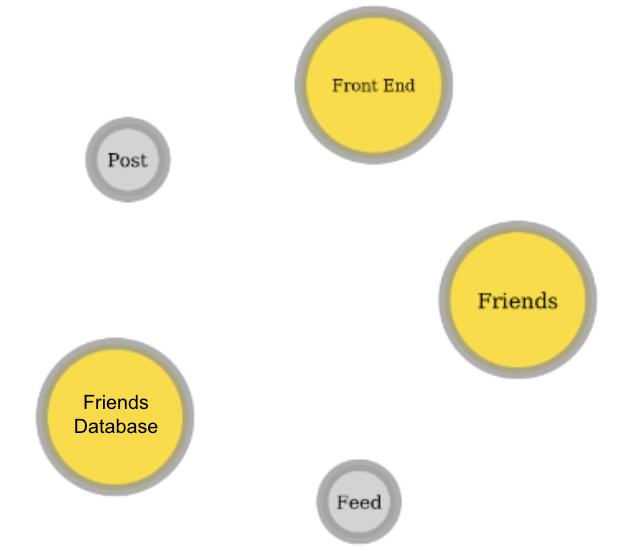}
    \minicaption{Single group visualization of Trace 1 and Trace 2}{}
    \label{fig:example_group_vis}
    \vspace{-0.5cm}
  \end{figure}
}

%% file: figures/fig_friends_group_vis.tex

{
\begin{figure}[t]
    \includegraphics[width=0.9\columnwidth]{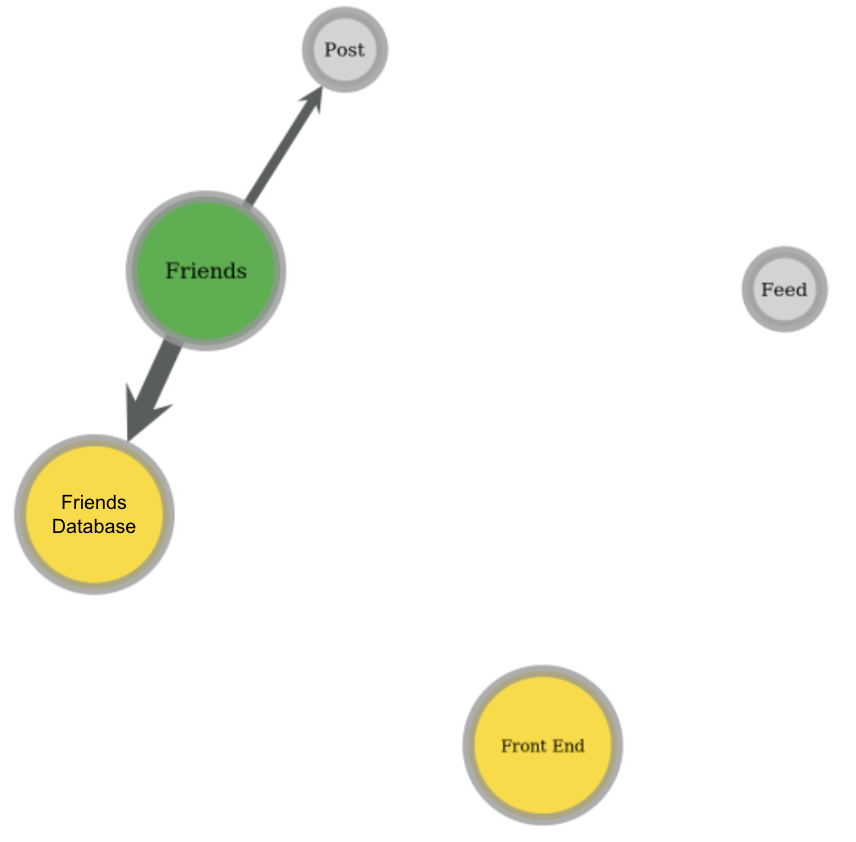}
    \minicaption{Chosen service visualization for Friends}{}
    \label{fig:example_friends_group}
    \vspace{-0.5cm}
  \end{figure}
}

%% file: sections/results.tex
\section{Results}
This is a brief overview of the results for our method of grouping similar traces. We investigate the outputs and performances of \textbf{Definition 1}, the first encoding method that we've looked at so far, and we find the complexity of our code. The Jaccardian Similarity calculated between traces was dependent on how many services they shared.

\subsection{Experimental Setup}
\subsubsection{Software Environment} 
We are using \href{https://www.python.org/downloads/release/python-3912/}{Python 3.9.12} for all of our programs. No configuration was required other than downloading Python. 

\subsubsection{Data for Experiment} 
JSON files are a common way of accessing data. JSON files are the text format of Javascript data structures. Most platforms like Jaeger and TrainTicket return their traces in the form of a JSON file so we used JSON files as input to our program to make our program more applicable to a majority of trace datasets. 

We wanted to test our method and program on a sample of 24 traces that we manually created. A pictorial representation of the sample traces is shown here (note that the traces are numbered 0 to 23 to keep it consistent with the program's indexing): \href{https://docs.google.com/presentation/d/18BaB7-JUgVcSynagCJTDJ1ZseoMkKPoDYU0umwd-mgM/edit?usp=sharing}{Sample Traces}. To apply this definition, which is based on the set of services in the trace, we listed the services in a JSON file excluding any repeats in the service names. For example, the JSON format for trace 4 is shown in Figure \ref{fig:example_json}. We also created a set of 32 traces by adding 8 other traces to our initial sample to add more variation to the list of service names (we added new service names to our trace system). 

\input{figures/fig_example_json_format}

JSON format of both traces can be found here: 

\href{https://drive.google.com/file/d/1P8MtXedjJd9YmrR9WdklKNUkhlJakK7N/view?usp=sharing}{trace1.json} - 24 traces sample dataset 

\href{https://drive.google.com/file/d/1TE1nSYpEcJzrCYBwaU7bMP6nLMuyIoKj/view?usp=sharing}{trace2.json} - 32 traces sample dataset

\subsubsection{Running Experiments} \
Our code is a single Python program so we run the program with the following command (Main.py is the file name): python3 Main.py. We ran two sets of experiments: 

\begin{itemize}
    \item Setting the threshold to our default value, 0.8 (don't apply the Binary Search method). We ran this set of experiments on trace1.json.
    \item Applying the optimal threshold finding method with the goal for the number of tracing groups as 6. We ran this set of experiments on trace2.json.
\end{itemize}

\subsection{Default Threshold Experiment}
\subsubsection{Results of Effectiveness} 
As shown in slide 5 on the sample traces page above, we manually applied trace definition 1 (categorization based on the names of the services). We listed the different possible arrays of service names as well as the traces that would be in that category, shown below: 
\begin{itemize}
    \item A,B,C,D: 0, 1, 3, 7, 8, 9, 11, 12, 21, 22
    \item A,B,C: 2, 5, 19
    \item A,B,D: 4, 13
    \item A,C,D: 6, 20, 23
    \item A,D: 10, 18
    \item B,D: 14, 15
    \item C,D: 16, 17
\end{itemize}
As a result of our program, we got groups of similar traces with the definition and one representative trace from each group that represents the entire group. The output for this set of sample traces is below: 

\includegraphics[width=\columnwidth]{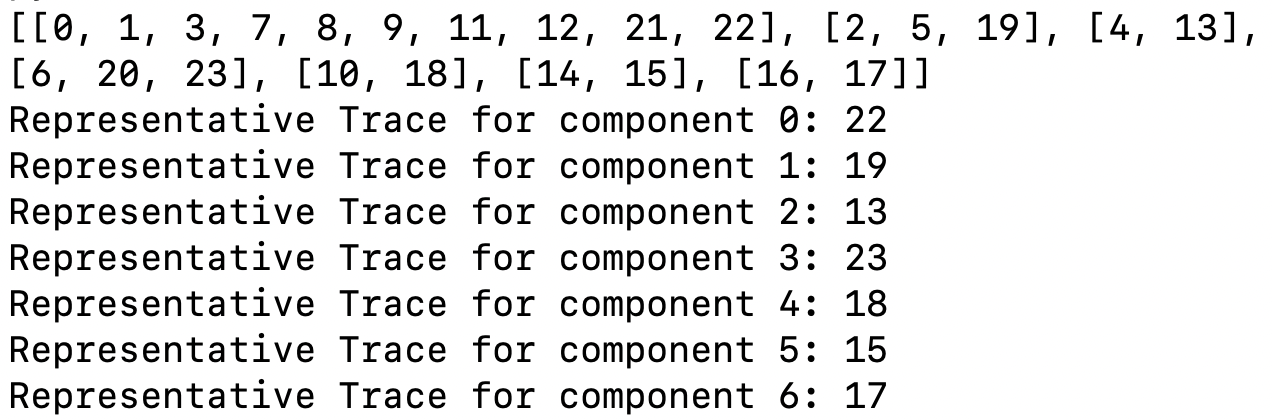} 

So, our grouping of traces exactly matched the grouping that we manually made. 

\subsubsection{Results of Performance} 
We used the time library in Python to measure the performance of our program. The runtime on average (over 4 runs of our program on the 24 sample trace data set) was 0.077 seconds. Our algorithm's time complexity is $O(n^2)$ but the time complexity for DSU can get much larger if the graph created using Jaccardian similarity has many edges. 

The exact time taken may vary whether the user is using a computer with more or less powerful hardware (but the time complexity won't change). 
\subsection{Optimal Threshold Experiment}
\subsubsection{Results of Effectiveness} 
We ran our method with the goal number of trace groups as 6 and then 11. 

As output for a goal of 6 trace groups we got: 

\includegraphics[width=\columnwidth]{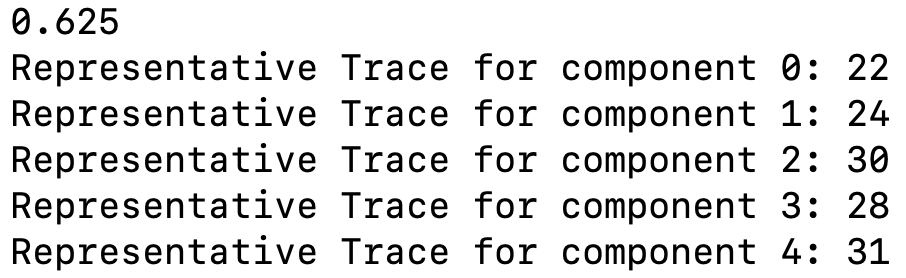} 

This shows that the optimal threshold is 0.625 and this threshold gets 5 trace groups. We manually checked our data and 5 trace groups is the closest we can get to 6 groups.

As output for a goal of 11 trace groups we got: 

\includegraphics[width=\columnwidth]{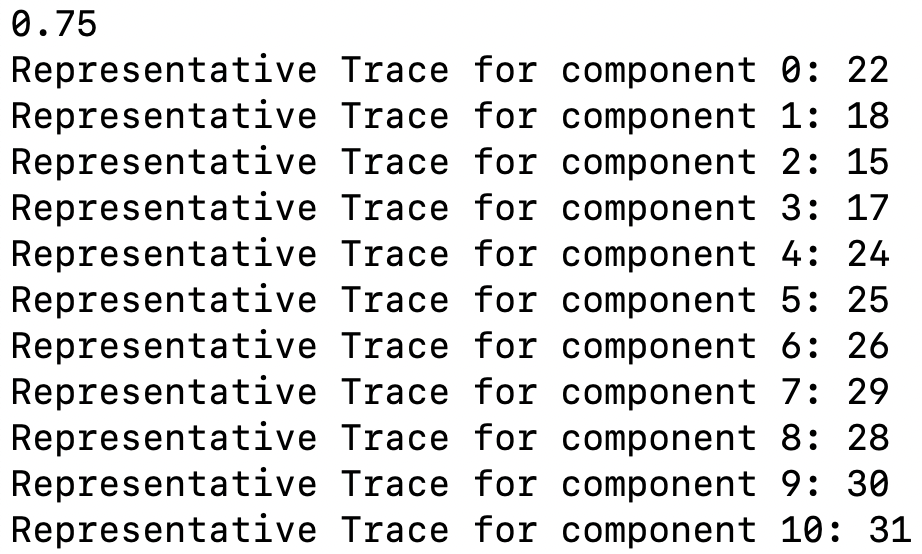} 

This shows that the optimal threshold is 0.75 and this threshold gets 11 trace groups.

So, we are able to find an optimal threshold value for a given set of traces and a goal number of trace groups.

\subsubsection{Results of Performance.} 
We still used the time library in Python to measure the performance of our program. Our algorithm's time complexity is $O(n^2 \log{n})$ but the time complexity for Binary Search and DSU can get much larger depending on the goal number of trace groups and the number of edges that the trace similarity graph has. For 6 trace groups, the program took an average of 8.476 seconds. For 11 trace groups, the program took an average of 0.506 seconds. 

The exact time taken may vary whether the user is using a computer with more or less powerful hardware (but the time complexity won't change).

%% file: figures/fig_example_json_format.tex
{
\begin{figure}[h!]
    \includegraphics[width=0.7\columnwidth]{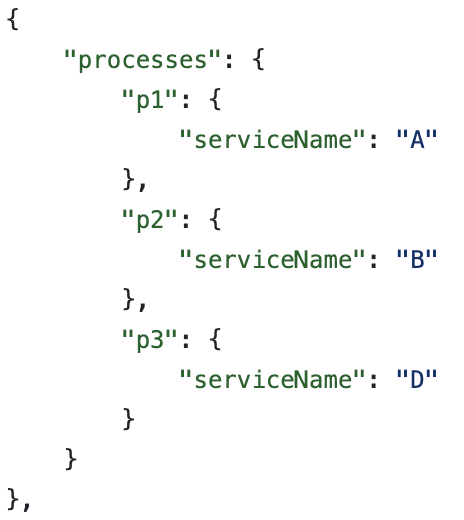}
    \minicaption{Example JSON format}{}
    \label{fig:example_json}
    \vspace{-0.5cm}
  \end{figure}
}

%% file: sections/future_work.tex
\section{Future Work}
\quad In the time ahead, we aim to investigate different ways to determine similarity between traces as well as optimal threshold values. Additionally, we will continue exploring possible visualizations of aggregate traces and create an algorithm to predict missing parts of traces.

We plan to test with datasets with larger amounts of traces to evaluate the performance of our definitions. We have hundreds of traces from DeathStarBench to look through. Once we see how our algorithm executes over them, we can make proper adjustments to our code or our definition. It will also give us a better idea of how to calculate an optimal similarity threshold.

One limitation of our method is that if one trace is a subgraph of another, it doesn't necessarily mean that the smaller trace is incomplete. Some workflows in a system terminate before other workflows in certain scenarios (like a cache hit vs miss) but it doesn't mean that the terminated trace is incomplete. To move forward with our current approach, we plan to look at the other details in the trace to have a better understanding as to whether a trace is actually an incomplete/complete version of another trace given that one is a subgraph of the other.

To move forward with this research question we want to look at different similarity measures. Our current method only implements trace encoding methods based on service names and trace structure but there are other unexplored variables in traces that could be used for measuring trace similarity such as, latency or request type. A disadvantage of our current approach to identifying and grouping similar traces is the time complexity, which would make it difficult to scale up significantly, especially if there are many services to consider. We also plan to look into ways we can improve the performance of our method.

In addition to our current visualizations of individual groups of traces, we also want to eventually implement a visualization of all groups within a dataset. We could distance the centers of these groups based on how similar they are depending on the definition we use so that developers can determine the most anomalous groups.

%% file: sections/conclusion.tex
\section{Conclusion}
\quad We were able to filter out incomplete traces if we found that they had a more complete version in the trace set. From this, we had a smaller and more informative set of traces that we used to analyze and visualize the traces. 

Using our first definition of trace similarity, we were able to implement a program that successfully groups similar traces based on the names of the services in the traces. We also used our third definition to group similar traces based on the exact structure of the trace graph. We analyzed our groups of similar traces to choose representative traces that would represent each of our groups, and we created visualizations for groups that highlights common services and common calls to services. Additionally, we were able to find the optimal threshold for a given set of traces and a goal for the number of trace groups. 

\section{Data and Source Code}
All of our data and source code are provided here: \url{https://github.com/docc-lab/PRIMES2023-VDTA}.